\documentclass[final,3p,times,twocolumn]{elsarticle}
\usepackage{amsmath,amssymb,graphicx}
\usepackage{amsfonts,amsthm}

\usepackage{amsmath,amssymb,graphicx,amsthm}
\usepackage{color}
\usepackage{soul}
\usepackage{epigraph}

\newcommand{\D}{{\mathrm{d}}}

\begin{document}

\begin{frontmatter}

\title{Detailed balance in micro- and macrokinetics \\ and micro-distinguishability of macro-processes}
\author{A.~N.~Gorban}
 \ead{ag153@le.ac.uk}
\address{Department of Mathematics, University of Leicester, Leicester, LE1 7RH, UK}

\date{}


\begin{abstract}
We develop a general framework for the discussion of detailed balance and analyse its microscopic background.
We find that there should be two additions to the well-known $T$- or $PT$-invariance of the microscopic laws of motion:

1. Equilibrium should not spontaneously break the relevant $T$- or $PT$-symmetry.

2. The macroscopic processes should be microscopically distinguishable to guarantee
persistence of detailed balance in the model reduction from micro- to macrokinetics.

We briefly discuss examples of the violation of these rules and the corresponding violation of detailed balance.
\end{abstract}

\end{frontmatter}

\section{The history of detailed balance in brief\label{sec1}}

\epigraph{VERY deep is the well of the past. ... For the deeper we sound, the further
down into the lower world of the past we probe and press, the more do we find that the
earliest foundation  of humanity, its history and culture, reveal themselves
unfathomable.}{T. Mann \cite{MannJoseph}}

\begin{figure}[h]
\centering{
\includegraphics[height=0.08\textwidth]{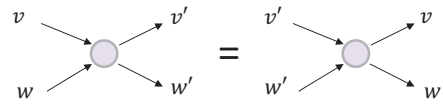}
}\caption{Schematic representation of detailed balance for collisions: at equilibrium, each collision is equilibrated by the reverse collision.
\label{scheme1}}
\end{figure}

Detailed balance as a consequence of the reversibility of collisions ({\em at
equilibrium, each collision is equilibrated by the reverse collision},
Fig.~\ref{scheme1}) was introduced by Boltzmann for the Boltzmann equation and used in
the proof of the $H$-theorem \cite{Boltzmann1872} (Boltzmann's arguments were analyzed by
Tolman \cite{Tolman1938}). Five years earlier, Maxwell used the principle of detailed
balance for gas kinetics with the reference to the {\em principle of sufficient reason}
\cite{Maxwell1867}. He analyzed equilibration in cycles of collisions and in the pairs of
mutually reverse collisions and mentioned ``Now it is impossible to assign a reason why
the successive velocities of a molecule should be arranged in this cycle, rather than in
the reverse order.''

In 1901, Wegscheider introduced detailed balance for chemical kinetics
on the basis of classical thermodynamics \cite{Wegscheider1901}.
He used the assumption that each elementary reaction is reversible and should respect thermodynamics
(i.e. entropy production in this reaction should be always non-negative). Onsager used this work of
Wegsheider in his famous paper \cite{Onsager1931}. Instead of direct citation he wrote: ``Here,
however, the chemists are accustomed to impose a very interesting
additional restriction, namely: when the equilibrium is reached each individual reaction must balance itself.''
Einstein used detailed balance as a basic assumption in his theory of radiation \cite{Einstein1916}.
In 1925, Lewis recognized the principle of detailed balance as a new general principle of equilibrium
\cite{Lewis1925}. The limit of the detailed balance for systems which include some irreversible
elementary processes (without reverse processes) was recently studied in detail \cite{GorbanYablonsky2011,GorbanMirkesYablonsky2013}.

In this paper, we develop a general formal framework for discussion of detailed balance, analyse its microscopic background and
persistence in the model reduction from micro- to macrokinetics.

\section{Sampling of events, $T$-invariance and detailed balance}

\subsection{How detailed balance follows from microreversibility}

In the sequel, we omit some technical details assuming that all the operations are
possible, all the distributions are regular and finite Borel (Radon) measures, and all
the integrals (sums) exist.

\begin{figure}[h]
\centering{
\includegraphics[width=0.3\textwidth]{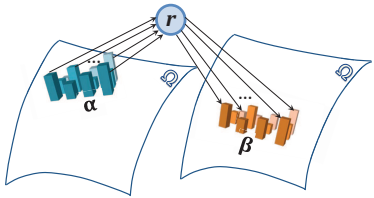}
}\caption{Schematic representation of an elementary process. Input ($\alpha$) and output ($\beta$) distributions are represented by column histograms.
\label{reactGen}}
\end{figure}

The basic notations and notions:
\begin{itemize}
\item $\Omega$ -- a space of states of a system (a locally compact metric space);
\item Ensemble $\nu$ -- a non-negative distribution on $\Omega$;
\item Elementary process has a form $\alpha \to \beta$ (Fig.~\ref{reactGen}), where $\alpha$, $\beta$
are non-negative distributions;
\item Complex -- an input or output distribution of an elementary process.
\item $\Upsilon$ -- the set of {\em all} complexes participating in elementary processes. It is equipped with the
weak topology and is a closed and locally compact set of distributions.
\item The reaction rate $r$ is a measure defined on $\Upsilon^2=\{(\alpha,\beta)\}$. It describes the rates of all elementary processes $\alpha \to \beta$.
\item  The support of $r$, supp$r\subset \Upsilon^2$, is the {\em mechanism} of the
    process, i.e. it is the set of pairs $(\alpha, \beta)$, each pair represents an
    elementary process $\alpha \to \beta$. (Usually, supp$r\varsubsetneq
    \Upsilon^2$.)
\item The rate of the whole kinetic process is a  distribution $W$ on $\Omega$ (the following integral should exist):
    $$W=\frac{1}{2}\int_{(\alpha,\beta)\in\Upsilon^2} (\beta-\alpha) \D [r(\alpha, \beta)-r(\beta,\alpha)].$$
\end{itemize}

The distribution $\nu$ depends on time $t$. For systems with continuous time,
$\dot{\nu}=W$. For systems with discrete time,  $\nu(t+\tau)-\nu(t)=W$, where $\tau$ is
the time step. To create the closed kinetic equation (the associated {\em nonlinear
Markov process} \cite{Kolokoltsov2010}) we have to define the map $\nu \mapsto r$ that
puts the reaction rate $r$ (a Radon measure on $\Upsilon^2$) in correspondence with a
non-negative distribution $\nu$ on $\Omega$ (the {\em closure problem}). In this
definition, some additional restrictions on $\nu$ may be needed. For example, one can
expect that $\nu$ is absolutely continuous with respect to a special (equilibrium)
measure. There are many standard examples of kinetic systems: mass action law for
chemical kinetics  \cite{Yablonskiiatal1991,MarinYablonsky2011}, stochastic models of
chemical kinetics \cite{Gillespie2007}, the Boltzmann equation \cite{Cercignani} in
quasichemical representation \cite{GorKarLNP2005} for space-uniform distributions, the
lattice Boltzmann models \cite{LB2}, which represent the space motion as elementary
discrete jumps (discrete time), and the quasichemical models of diffusion
\cite{GorbanSarkisyan2011}.

We consider interrelations between two important properties of the measure $r(\alpha, \beta)$:

({EQ}) $W=0$ (equilibrium condition);

({DB}) $r(\alpha, \beta)\equiv r(\beta,\alpha)$ (detailed balance condition).

It is possible to avoid the difficult closure question about the map $\nu \mapsto r$ in
discussion of $T$-invariance and relations between EQ and DB conditions.

Obviously, DB$\Rightarrow$EQ. There exists a trivial case when  EQ$\Rightarrow$DB (a sort
of linear independence of the vectors $\gamma=\beta-\alpha$ for elementary processes
joined in pairs with their reverse processes): if ($\mu(\alpha,
\beta)=-\mu(\beta,\alpha)$) $$\int_{(\alpha,\beta)\in {\rm supp}r}(\beta-\alpha) \D
\mu(\alpha,\beta)=0 \Rightarrow \mu=0$$ for every antisymmetric measure $\mu$ on
$\Upsilon^2$ ($\mu(\alpha, \beta)=-\mu(\beta,\alpha)$), then EQ$\Rightarrow$DB.

There is a much more general reason for detailed balance, $T$-invariance. Assume that the
kinetics give a coarse-grained description of an ensemble of interacting microsystems and
this interaction of microsystems obeys a reversible in time equation: if we look on the
dynamics backward in time  ({\em operation $T$}) we will observe the solution of the same
dynamic equations. For $T$-invariant microscopic dynamics, $T$ maps an equilibrium
ensemble into an equilibrium ensemble. Assuming uniqueness of the equilibrium under given
values of the conservation laws, one can just postulate the {\em invariance of equilibria
with respect to the time reversal transformation} or $T$-invariance of equilibria: if we
observe an equilibrium ensemble backward in time, nothing will change.

Let the complexes remain unchanged under the action of $T$. In this case, the time
reversal transformation for collisions (Fig.~\ref{scheme1}) leads to the reversal of
arrow: the direct collision is transformed into the reverse collision. The same
observation is valid for inelastic collisions. Following this hint, we can accept that
the reversal of time $T$ transforms every elementary process $\alpha \to \beta$ into its
reverse process $\beta \to \alpha$. This can be considered as a restriction on the
definition of direct and reverse processes in the modelling (a ``model engineering''
restriction): the direct process is an ensemble of microscopic events and the reverse
process is the ensemble of the time reversed events.

Under this assumption, $T$ transforms $r(\alpha,\beta)$ into $r(\beta,\alpha)$. If the
rates of elementary processes may be observed (for example, by the counting of
microscopic events in the ensemble) then $T$-invariance of equilibrium gives DB: at
equilibrium, $r(\alpha,\beta)=r(\beta,\alpha)$, i.e. EQ$\Rightarrow$DB under the
hypothesis of $T$-in\-va\-riance.

The assumption that the complexes are invariant under the action of $T$ may be violated:
for example, in Boltzmann's collisions (Fig.~\ref{reactGen}) the input measure is
$\alpha=\delta_{v}+\delta_{w}$ and the output measure is $\beta=\delta_{v'}+\delta_{w'}$.
Under time reversal, $\delta_{v}{ \overset{T}\mapsto} \delta_{-v}$. Therefore $\alpha{
\overset{T}\mapsto} \delta_{-v}+\delta_{-w}$ and $\beta { \overset{T}\mapsto}
\delta_{-v'}+\delta_{-w'}$. We need an additional invariance, the space inversion
invariance (transformation $P$) to prove the detailed balance (Fig.~\ref{scheme1}).
Therefore, the detailed balance condition for the Boltzmann equation (Fig.~\ref{scheme1})
follows not from $T$-invariance alone but from $PT$-invariance because for Boltzmann's
kinetics $$\{\alpha \to \beta\} { \overset{PT}\mapsto} \{\beta \to \alpha\}.$$

In any case, the microscopic reasons for the detailed balance condition include existence
of a symmetry transformation $\mathfrak{T}$ such that
\begin{equation}\label{Tdirecttoreverse}
\{\alpha \to \beta\} { \overset{\mathfrak{T}}\mapsto} \{\beta \to \alpha\}
\end{equation}
and the microscopic dynamics is invariant with respect to  $\mathfrak{T}$. In this case,
one can conclude that (i) the equilibrium is transformed by $\mathfrak{T}$ into the same
equilibrium (it is, presumably, unique) and (ii) the reaction rate $r(\alpha, \beta)$ is
transformed into $r(\beta,\alpha)$ and does not change because nothing observable can
change (equilibrium is the same). Finally, at equilibrium $r(\alpha, \beta)\equiv
r(\beta,\alpha)$ and EQ$\Rightarrow$DB.

There remain two question:
\begin{enumerate}
\item We are sure that $\mathfrak{T}$ transforms the equilibrium state into an
    equilibrium state but is it necessarily the same equilibrium? Is it forbidden
    that the equilibrium is degenerate and $\mathfrak{T}$ acts non-trivially on the
    set of equilibria?
\item We assume that the rates of different elementary processes are physical
    observables and the ensemble with different values of these rates may be
    distinguished experimentally. Is it always true?
\end{enumerate}

The answer to both questions is ``no''. The principle of detailed balance can be violated
even if the physical laws are $T$, $P$ and $PT$ symmetric. Let us discuss the possible
reasons for these negative answers and the possible violations of detailed balance.

\subsection{Spontaneous breaking  of $\mathfrak{T}$-symmetry}
Spontaneous symmetry breaking is a well known effect in phase transitions and particle
physics. It appears when the physical laws are invariant under a transformation, but the equilibrium of the
system transforms into another state, which should be also equilibrium. Hence, the equilibrium is degenerated.
The best known examples are magnets. They are not rotationally symmetric (there is a continuum of equilibria that
differ by the direction of magnetic field). Crystals are not symmetric with respect to
translation (there is continuum of equilibria that differ by a shift in space). In these
two examples,  the multiplicity of equilibria is masked by the fact that all these
equilibrium states can be transformed into each other by a proper rigid motion
transformation (translation and rotation).

The {\em nonreciprocal media} violate $T$ and  $PT$ invariance
\cite{Krowne1993,Kamenetskii1998,GuoetAl2009}. These media are transformed by $T$ and $PT$ into
different ({\em dual}) equilibrium media and cannot be transformed back by a proper rigid
motion. The implication EQ$\Rightarrow$DB for the nonreciprocal media may be
wrong and for its validity some strong additional assumptions are needed, like the linear
independence of elementary processes.

Spontaneous breaking of $\mathfrak{T}$-symmetry provides us a counterexample to the proof
of detailed balance. In this proof, we used the assumption that under transformation
$\mathfrak{T}$ elementary processes transform into their reverse processes
(\ref{Tdirecttoreverse}) and, at the same time, the equilibrium ensemble does not change. 

If the equilibrium is transformed by $\mathfrak{T}$ into another (but obviously also
equilibrium) state then our reasoning cannot be applied to reality and the proof is not
valid. Nevertheless, the refutation of the proof does not mean that the conclusion
(detailed balance) is necessarily  wrong. Following the Lakatos  terminology \cite{Lakatos}
we should call the spontaneous breaking of $\mathfrak{T}$-symmetry the {\em local
counterexample} to the principle detailed balance. It is an intriguing question whether
such a local counterexample may be transformed into a {\em global} one: does the
violation of the Onsager reciprocal relation mean the violation of detailed balance (and
not only the refutation of its proof)?

\subsection{Reciprocal relation and detailed balance}

It is known that for many practically important kinetic laws the Onsager reciprocal
relations follow from detailed balance. In these cases, violation of the reciprocal
relations implies violation of the principle of detailed balance. For example, for the systems
in magnetic fields the reciprocal relations may be violated \cite{DeGrootMazur}, and we can
expect that detailed balance for these systems will be also violated.

For master equation (first order kinetics or continuous time Markov chains) the principle of detailed balance
is {\em equivalent} to the reciprocal relations (\cite{DeGrootMazur} Ch. 10, $\S$ 4). For the
nonlinear mass action law the implication ``detailed balance $\Rightarrow$ reciprocal
relations'' is also well known (see, for example, \cite{Yablonskiiatal1991}) but the
equivalence is not correct because the number of nonlinear reactions for a given number
of components may be arbitrarily large and it is possible to select such values of
reaction rate constants that the reciprocal relations are satisfied but the principle of
detailed balance does not hold. For transport processes, the quasichemical models
\cite{GorbanSarkisyan2011} also demonstrate how the reciprocal relations follow from
detailed balance for the mass action law kinetics or the generalized mass action law. 
We confine the discussion of kinetic laws to systems with finite sets of components.

Consider a finite-dimensional system with the set of components (species or states)
$A_1,\ldots, A_n$ given. For each $A_i$ the extensive variable $N_i$ (``amount" of
$A_i$) is defined. The  Massieu–-Planck function $\Phi(N,\ldots)$ ({\em free entropy}
\cite{Callen1985}) depends on the vector $N$ with coordinates $N_i$ and on the variables
that are constant under given conditions. For isolated systems instead of $(\ldots)$ in
$\Phi$ we should use internal energy $U$ and volume $V$ (and this $\Phi$ is the entropy),
for isothermal isochoric systems these variables are $1/T$ and $V$, where $T$ is
temperature, and for isothermal isobaric systems we should use $1/T$ and $P/T$, where $P$
is pressure. For all such conditions, $$\frac{\partial \Phi}{\partial
N_i}=-\frac{\mu_i}{T},$$ where $\mu_i$ is the chemical potential of $A_i$ or the {\em
generalized chemical potential} for the quasichemical models where interpretation of
$A_i$ is wider than just various atomic particles.

Elementary processes in the finite-dimensional systems are represented by their
stoichiometric equations $$\sum_i \alpha_{{\rho}i} A_i \to \sum_i \beta_{{\rho}i} A_i.$$
This is a particular case of the general picture presented in Fig.~\ref{reactGen}. The
{\em stoichiometric vector} is $\gamma_{\rho}$:
$\gamma_{{\rho}i}=\beta_{{\rho}i}-\alpha_{{\rho}i}$ (gain minus loss). The generalized
mass action law represents the reaction rate in the following form:
\begin{equation}\label{gemMAL}
 r_{\rho}=\phi_{\rho} \exp\left(\sum_i \alpha_{{\rho}i}\frac{\mu_i}{RT}\right),
\end{equation}
where $\exp(\sum_i \alpha_{{\rho}i}{\mu_i}/{RT})$ is the Boltzmann factor ($R$ is the gas
constant) and $\phi_r>0$ is the kinetic factor (this representation is closely related to
the transition state theory \cite{Eyring1935} and its generalizations
\cite{GorbanShach2011}).

The equilibria and conditional equilibria are described as the maximizers of the free
entropy under given conditions. For a system with detailed balance every elementary
process has a reverse process and the couple of processes $\sum_i \alpha_{{\rho}i} A_i
\rightleftharpoons \sum_i \beta_{{\rho}i} A_i$ should move the system from the initial
state to the partial equilibrium, that is the maximizer of the function $\Phi$ in the
direction $\gamma_{\rho}$. Assume that the equilibrium is not a {\em boundary point} of
the state space. For a smooth function $\Phi$, the conditional maximizer in the direction
$\gamma_{\rho}$ should satisfy the necessary condition $\sum_i \gamma_{{\rho}i}\mu_i=0$.
In the generalized mass action form (\ref{gemMAL}) the detailed balance condition has a
very simple form:
\begin{equation}\label{DetBalGMAL}
\phi_{\rho}^+=\phi_{\rho}^-,
\end{equation}
where $\phi_{\rho}^+$ is the kinetic factor for the direct reaction and $\phi_{\rho}^-$
is the kinetic factor for the reverse reaction.

Assume that the detailed balance condition (\ref{DetBalGMAL}) holds. Let us join the elementary processes in pairs,  direct with reverse ones, with the corresponding change in their numeration. The kinetic equation is $\dot{N}=V\sum_{\rho}
\gamma_{\rho} (r_{\rho}^+-r_{\rho}^-)$. The Jacobian matrix at equilibrium is
$$\left.\frac{\partial\dot{N}_i}{\partial N_j}\right|_{\rm eq} =
-\frac{V}{R} \sum_k \left( \sum_{\rho} r_{\rho}^{\rm eq} \gamma_{{\rho}i}\gamma_{{\rho}k}\right) \left.\frac{\partial
(\mu_k/T)}{\partial N_j}\right|_{\rm eq},$$
 where $r_{\rho}^{\rm eq}=r_{\rho}^{+{\rm eq}}=r_{\rho}^{-{\rm eq}}$ is the rate at equilibrium of the
 direct and reverse reactions (they coincide due to detailed balance) and the subscript
 `eq' corresponds to the derivatives at the equilibrium. The linear approximation to the
 kinetic equations near the equilibrium is
 $$\frac{\D {\Delta N_i}}{\D t}=-\frac{V}{R}  \sum_k \left(\sum_{\rho} r_{\rho}^{\rm eq} \gamma_{{\rho}i}\gamma_{{\rho}k}\right) \Delta
 \left(\frac{\mu_k}{T}\right),$$
where $\Delta N_i$  and $\Delta(\mu_k/T)$  are deviations from the equilibrium values.
The variables $\Delta N_i$  are extensive thermodynamic coordinates and  $\Delta(\mu_k/T)$ are intensive conjugated
variables -- thermodynamic forces. Time derivatives $\D {\Delta N_i}/ \D t$ are thermodynamic fluxes.
Symmetry of the matrix of coefficients and, therefore, validity of the reciprocal relations is obvious.

Thus, for a wide class of kinetic laws the reciprocal relations in a vicinity of a
regular (non-boundary) equilibrium point follow from detailed balance in the linear
approximation. In these cases, the non-reciprocal media give {\em global
counterexamples} to the detailed balance. Without reference to a kinetic law they
remain local counterexamples to the proof of detailed balance.

\subsection{Sampling of different macro-events from the same micro-events}
In kinetics, only the total rate $W$ is observable (as $W=\dot{\nu}$ or $W= \Delta \nu =
\nu (t+\tau)-\nu(t)$). In the macroscopic world the observability of the  rates of the
elementary processes is just a hypothesis.

Imagine a microscopic demon that counts collisions or other microscopic events of various
types. If different elementary processes correspond to different types of microscopic
events then the rates of elementary processes can be observed. If the equilibrium
ensemble is invariant with respect to $\mathfrak{T}$ then the demon cannot detect the
difference between the equilibrium and the transformed equilibrium and the rates of
elementary processes should satisfy DB. But it is possible to sample the elementary
processes of macroscopic kinetics from the events of microscopic kinetics in different
manner.

For example, in chemical mass action law kinetics we can consider the reaction mechanism
$A\rightleftharpoons B$ (rate constants $k_{\pm 1}$), $A+B\rightleftharpoons 2B$ (rate
constants $k_{\pm 1}$) \cite{Joshi2013}. We can also create a stochastic model for this
system with the states $(x A,y B)$ ($x$, $y$ are nonnegative integers) and the elementary
transitions $(x A,y B)\rightleftharpoons ((x-1) A, (y+1) B)$ (rate constants
$\kappa_+=k_{+1}x+k_{+2}x^2$, $\kappa_-=k_{-1}(y+1)+k_{-2}(x-1)(y+1)$). The elementary
transitions in this stochastic model are linearly independent and EQ$\Leftrightarrow$DB.
In the corresponding mass action law chemical kinetics detailed balance requires
additional relation between constants: $k_{+1}/k_{-1}=k_{+2}/k_{-2}$.

Thus, macroscopic detailed balance may be violated in this example when microscopic
detailed balance holds. (For more examples and theoretic consideration of the relations
between detailed balance in mass action law chemical kinetics and stochastic models of
these systems see \cite{Joshi2013}.) Indeed, both of the macroscopic elementary processes
$A\rightleftharpoons B$ and $A+B\rightleftharpoons 2B$ correspond to the same set of
microscopic elementary processes $(x A,y B)\rightleftharpoons ((x-1) A, (y+1) B)$. Each
of these elementary event is ``shared'' between two different macroscopic elementary
processes. Therefore, the macroscopic elementary processes in this example are  {\em
microscopically indistinguishable}.

The microscopic indistinguishability in this example follows from the coincidence of the
stoichiometric vectors for two macroscopic processes $A \rightleftharpoons B$ and
$A+B\rightleftharpoons 2B$. If the stoichiometric vectors are just linear dependent then
it does not imply microscopic indistinguishability.

For example, let us take two
reactions $A\rightleftharpoons B$ and $2A \rightleftharpoons 2B$. For the first reaction
the corresponding microscopic processes have the form $(x A,y B)\rightleftharpoons ((x-1)
A, (y+1) B)$ (if all the coefficients are nonnegative). For the reaction $2A
\rightleftharpoons 2B$ the microscopic processes have the form $(x A,y
B)\rightleftharpoons ((x-2) A, (y+2) B)$ (if all the coefficients are nonnegative). These
sets do not intersect, the elementary processes are microscopically distinguishable and
macroscopic detailed balance follows from microscopic detailed balance.

Nontrivial Wegscheider identities appear in this example  at the
microscopic level (in the first example all the microscopic transitions are linearly
independent and there exist no additional relations). Let the microscopic reaction rate
constants for the reaction $(x A,y B)\rightleftharpoons ((x-1) A, (y+1) B)$ be
$\kappa_1^{\pm}(x,y)$ and $\kappa_2^{\pm}(x,y)$ for the reaction $(x A,y
B)\rightleftharpoons ((x-2) A, (y+2) B)$. Due to detailed balance,  in each cycle of
a linear reaction network  the product of reaction rate constants in the clockwise
direction coincides with the product in the anticlockwise direction. It is sufficient
to consider the basis cycles (and their reversals):
\begin{equation*}
 \begin{split} (x A,y B)& \to ((x-1) A, (y+1) B)\to \\  & \to ((x-2) A, (y+2) B) \to (x A,y B).\end{split}
\end{equation*}
Therefore,
\begin{equation*}
\begin{split}
\kappa_1^{+}(x,y)&\kappa_1^{+}(x-1,y+1)\kappa_2^{-}(x,y) \\ &=
\kappa_2^{+}(x,y)\kappa_1^{-}(x-1,y+1)\kappa_1^{-}(x,y).
\end{split}
\end{equation*}
In the macroscopic limit these conditions transform into the macroscopic detailed balance
conditions.

\section{Relations between elementary processes beyond microreversibility and detailed balance}

If microreversibility does not exist, is everything permitted? What are the the relations
between the reaction rates beyond the microreversibility conditions if such universal
relations exist? The radical point of view is: beyond the microreversibility we face just
the world of kinetic equations with preservation of positivity, various specific
restrictions on the coefficients appear in some specific cases and the variety of these
cases in unobservable. Development of this point of view leads to the general theory of
nonlinear Markov processes \cite{Kolokoltsov2010}, i.e. the general theory of kinetic
equations with preservation of positivity.

\begin{figure}[t]
\centering{
\includegraphics[height=0.08 \textwidth]{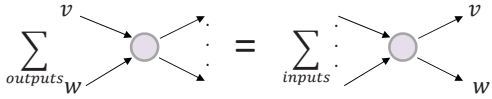}
}\caption{\label{ChemeComBal}Boltzmann's  cyclic  balance (1887) (or semi-detailed balance or complex balance) is a
summarised detailed balance condition: at equilibrium the sum of intensities of collisions with a given
input $v+w\to \ldots$ coincides with the sum of intensities of collisions with the same output $\ldots \to v+w$ (for general systems see (\ref{complBal})).}
\end{figure}

The problem of the relations between elementary processes beyond microreversibility and
detailed balance was stated by Lorentz in 1887 \cite{Lorentz1887}. Boltzmann immediately
proposed the solution \cite{Boltzmann1887} and used it for extension of his $H$-theorem
beyond microreversibility. These conditions have the form of partially summed
conditions of detailed balance (Fig.~\ref{ChemeComBal}, compare to Fig.~\ref{scheme1}).
This solution was analyzed, generalized and proved by several generations of researchers
(Heitler, Coester, Watanabe, Stueckelberg \cite{Stueckelberg1952} and others, see the review in
\cite{GorbanShach2011}). It was rediscovered in 1972 \cite{HornJackson1972} in the
context of chemical kinetics and popularized as the {\em complex balance condition}.

For finite-dimensional systems which obey the generalized mass action  law
(\ref{gemMAL}) the complex balance condition is also the summarized detailed balance
condition (\ref{DetBalGMAL}). Consider the set $\Upsilon$ of all input and output vectors
$\alpha_{\rho}$ and $\beta_{\rho}$. The complex balance condition reads: for every $y\in
\Upsilon$
\begin{equation}\label{complBal}
\sum_{\rho, \, \alpha_{\rho}=y} \phi_{\rho}=\sum_{\rho,\, \beta_{\rho}=y} \phi_{\rho}.
\end{equation}

\begin{figure}
\centering{
\includegraphics[width=0.4\textwidth]{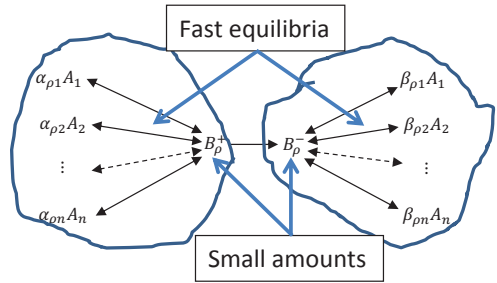}}
\caption{\label{StueckAsympt}Schematic representation of the Michaelis--Menten--Stueckelberg asymptotic
assumptions: an elementary process $\sum \alpha_{\rho i}A_i \to \sum \alpha_{\rho i}A_i$ goes through intermediate compounds $B_{\rho}^{\pm}$. The fast equilibria
$\sum \alpha_{\rho i}A_i\rightleftharpoons B_{\rho}^+$ and $\sum \beta_{\rho i}A_i\rightleftharpoons B_{\rho}^-$
can be described by conditional maximum of the free entropy. Concentrations of $B_{\rho}^{\pm}$ are small and reaction
between them obeys  linear kinetic equation.}
\end{figure}

Now, the complex balance conditions in combination with generalized mass action law are
proven for the finite-dimensional systems in the asymptotic limit proposed first by
Michaelis and Menten \cite{MichaelisMenten1913} for fermentative reactions and
Stueckelberg \cite{Stueckelberg1952} for the Boltzmann equation. This limit is
constituted by three assumptions (Fig.~\ref{StueckAsympt}): (i) the elementary processes
go through the intermediate compounds, (ii) the compounds are in fast equilibria with the
components (therefore, these equilibria can be described by thermodynamics) and (iii) the
concentrations of compounds are small with respect to concentrations of components
(hence, (iiiA) the quasi steady state assumption is valid for the compound kinetics and
(iiiB) the transitions between compounds follow the first order kinetics)
\cite{GorbanShach2011}. (It is worth mentioning that Michaelis and Menten in 1913 \cite{MichaelisMenten1913}
found the asymptotic limit  where the fermentative reaction can be described by the mass action
law. The so-called Michaelis--Menten kinetics is different and was invented 12 years later by Haldane and Briggs
\cite{BriggsHaldane1925}).

Thus, beyond microreversibility, Boltzmann's cyclic balance (or semi-detailed
balance, or complex balance) holds and it is as universal as the idea of intermediate
compounds (activated complexes or transition states) which exist in small concentrations
and are in fast equilibria with the basic reagents.

\section{Conclusion}

Thus, EQ$\Leftrightarrow$DB if:
\begin{enumerate}
\item{There exists a transformation $\mathfrak{T}$ that transforms the elementary
    processes into reverse processes and the microscopic laws of motion are
    $\mathfrak{T}$-invariant;}
\item{The equilibrium is symmetric with respect to $\mathfrak{T}$, that is, there is
    no spontaneous breaking of $\mathfrak{T}$-symmetry;}
\item{The macroscopic elementary processes are microscopically distinguishable. That
    is, they represent disjoint sets of microscopic events.}
\end{enumerate}
In applications, $\mathfrak{T}$ is usually either time reversal $T$ or the combined
transform $PT$.

For  level jumping (reduction of kinetic models \cite{GorKarLNP2005}), the equivalence
EQ$\Leftrightarrow$DB persists in the reduced (``macroscopic'') model if:
\begin{enumerate}
\item EQ$\Leftrightarrow$DB in the original (``microscopic'') model;
\item Equilibria of the macroscopic model correspond to  equilibria of the
    microscopic model. That is, the reduced kinetic model has no equilibria, which
    correspond to non-stationary dynamical regimes of the original kinetic model;
\item The macroscopic elementary processes are microscopically distinguishable. That
    is, they represent disjoint sets of microscopic processes.
\end{enumerate}

In this note, we avoid the discussion of an important part of Boltzmann's legacy which is
very relevant to the topic under consideration. Boltzmann represented kinetic process as
an {\em ensemble of indivisible elementary events --- collisions}. In the microscopic
world, a collision is a continuous in time and infinitely divisible process (and it
requires infinite time in most of the models of pair interaction). In the macroscopic
world it is instant and indivisible. The transition from continuous motion of particles
to an ensemble of indivisible instant collisions is not digested by modern mathematics up
to now, more than 130 years after its invention. The known results
\cite{Lanford1975,GallagherS-R} state that the Boltzmann equation for an ensemble of
classical particles with pair interaction and short--range potentials is asymptotically
valid starting from a non-correlated state during a fraction of the mean free flight
time. That is very far from the area of application. Nevertheless, if we just accept that
it is possible to count microscopic events then the reasons of validity and violations of
detailed balance in kinetics are clear.

\end{document}